\documentclass[article,nojss,shortnames]{jss}
\let\Prob\relax

\let\pkg\relax
\let\proglang\relax

\usepackage{thumbpdf}
\usepackage{amsfonts,amstext,amsmath,amssymb,amsthm}
\usepackage{accents}
\usepackage{color}
\usepackage{verbatim}
\usepackage{xspace}
\usepackage{booktabs}

\usepackage{fancyhdr}
\usepackage{newtxtext, newtxmath} 

\usepackage[utf8]{inputenc}
\usepackage{booktabs} \usepackage{amsfonts,amsmath}\usepackage{bm}                       \usepackage[authoryear]{natbib}                 \setcitestyle{aysep={}} \bibpunct{(}{)}{,}{a}{}{,} \usepackage{url}

\usepackage{hyperref}  

\definecolor{Red}{rgb}{0.5,0,0}
\definecolor{Blue}{rgb}{0,0,0.5}

 \hypersetup{hyperindex = {true},
    colorlinks = {true},
    linktocpage = {true},
    plainpages = {false},
    linkcolor = {Blue},
    citecolor = {Blue},
    urlcolor = {Red},
    pdfstartview = {Fit},
    pdfview = {XYZ null null null}
  }

\usepackage{numprint}
\npthousandsep{'}
\npdecimalsign{.}

\usepackage{setspace}
\usepackage{caption}
\usepackage{booktabs}

\newcommand{\rY}{Y}
\newcommand{\rYY}{\mY}
\newcommand{\rX}{\mX}

\newcommand{\ry}{y}
\newcommand{\rx}{\xvec}

\newcommand{\h}{h}

\newcommand{\parm}{\varthetavec}

\newcommand{\shiftparm}{\betavec}

\newcommand{\ie}{i.e.,~}
\newcommand{\eg}{e.g.,~}

\newcommand{\Prob}{\mathbb{P}}
\newcommand{\Ex}{\mathbb{E}}

\newcommand{\RR}{\mathbb{R}}
\newcommand{\NN}{\mathbb{N}}
\newcommand{\eps}{\varepsilon}

\usepackage{dsfont}

 \DeclareMathOperator{\diag}{diag}
 
 \DeclareMathOperator{\df}{df}

 \DeclareMathOperator{\ND}{N}
 
 \DeclareMathOperator{\UD}{U}

\def \avec {{\boldsymbol{a}}}

\def \hvec {{\boldsymbol{h}}}

    \def \mU {\boldsymbol{U}}

\def \xvec {\boldsymbol{x}}    \def \mX {\boldsymbol{X}}
\def \yvec {\boldsymbol{y}}    \def \mY {\boldsymbol{Y}}
\def \zvec {\boldsymbol{z}}

\def \betavec         {\boldsymbol{\upbeta}}\def \gammavec        {\boldsymbol{\upgamma}}

\def \thetavec        {\boldsymbol{\uptheta}}
\def \varthetavec     {\boldsymbol{\upvartheta}}

\def \xivec           {\boldsymbol{\upxi}}

\def \psivec          {\boldsymbol{\uppsi}}

\def \mLambda  {\boldsymbol{\Lambda}}

\def \mSigma   {\boldsymbol{\Sigma}}

\def \nullvec {\boldsymbol{0}}
\def \onevec {\boldsymbol{1}}

\usepackage[]{graphicx}\usepackage[]{xcolor}
\makeatletter
\def\maxwidth{ \ifdim\Gin@nat@width>\linewidth
    \linewidth
  \else
    \Gin@nat@width
  \fi
}
\makeatother

\definecolor{fgcolor}{rgb}{0.345, 0.345, 0.345}
\newcommand{\hlsng}[1]{\textcolor[rgb]{0.192,0.494,0.8}{#1}}\newcommand{\hldef}[1]{\textcolor[rgb]{0.345,0.345,0.345}{#1}}\newcommand{\hlkwc}[1]{\textcolor[rgb]{0.333,0.667,0.333}{#1}}\newcommand{\hlkwd}[1]{\textcolor[rgb]{0.737,0.353,0.396}{\textbf{#1}}}

\usepackage{framed}
\makeatletter
\newenvironment{kframe}{\def\at@end@of@kframe{}\ifinner\ifhmode \def\at@end@of@kframe{\end{minipage}}\begin{minipage}{\columnwidth}\fi\fi \def\FrameCommand##1{\hskip\@totalleftmargin \hskip-\fboxsep
 \colorbox{shadecolor}{##1}\hskip-\fboxsep
\hskip-\linewidth \hskip-\@totalleftmargin \hskip\columnwidth}\MakeFramed {\advance\hsize-\width
   \@totalleftmargin\z@ \linewidth\hsize
   \@setminipage}}{\par\unskip\endMakeFramed \at@end@of@kframe}
\makeatother

\definecolor{shadecolor}{rgb}{.97, .97, .97}
\definecolor{messagecolor}{rgb}{0, 0, 0}
\definecolor{warningcolor}{rgb}{1, 0, 1}
\definecolor{errorcolor}{rgb}{1, 0, 0}
\newenvironment{knitrout}{}{} 

\usepackage{alltt}
\IfFileExists{upquote.sty}{\usepackage{upquote}}{}
 
\usepackage{paralist}

\renewcommand{\thefootnote}{}

\newcommand{\snote}[1]{\ifdefined\draft \sidenote{#1}\fi}

\newcommand{\pkg}[1]{\textbf{#1}}
\newcommand{\proglang}[1]{\textsf{#1}}

\usepackage{moreverb} \immediate\write18{texcount -inc -incbib -sum -total bird_counts.tex > /tmp/wordcount.tex}

\newcommand{\mykeywords}{abundance data;
  count data;
  covariate-dependent Gaussian copula;
  distribution-free models;
  Gaussian copula;
  interspecific correlation;
  joint species distribution models; 
  transformation models
}

\title{Joint count transformation models with covariate-dependent correlations}

\author{
  Lukas Graz$^*$ \\ Universit\"at Z\"urich \And
  Luisa Barbanti$^*$\\ Universit\"at Z\"urich \And
  Roland Brandl \\ Universit\"at Marburg \And
  Torsten Hothorn \\ Universit\"at Z\"urich
}
\Plainauthor{Graz, Barbanti, Brandl and Hothorn
}
\Plaintitle{Joint count transformation models with covariate-dependent correlations}
\Shorttitle{Joint count transformation models with covariate-dependent correlations}

\Address{
  Lukas Graz, Luisa Barbanti, Torsten Hothorn\\
  torsten.hothorn@R-project.org\\
  Institut f\"ur Epidemiologie, Biostatistik und Pr\"avention\\
  Universit\"at Z\"urich\\
  Hirschengraben 84\\
  CH-8001 Z\"urich\\
  Switzerland

  Roland Brandl \\
Philipps-Universit\"at Marburg\\
  Karl-von-Frisch-Stra{\ss}e 8\\ DE-35032 Marburg\\ Germany
}

\Abstract{

Joint Species Distribution Models are essential for understanding how ecological covariates shape species communities. However, most existing approaches are limited by rigid parametric distributions for count data and the inability to model how interspecific associations change with those covariates. We introduce joint count transformation models, a novel framework designed to overcome these limitations.
Our approach combines distribution-free marginal count transformation models for multiple species with a covariate-dependent latent Gaussian copula to model interspecific correlations, interpretable as Spearman's rank correlation on the observed count scale. All model parameters are estimated efficiently via joint maximum likelihood estimation, implemented in the \proglang{R} package \pkg{tram}.

We apply this framework to model the joint abundance of three fish-eating bird species, using seasonality as the primary covariate. Our model successfully captured the complex, species-specific seasonal abundance patterns, including periods of high zero-counts and seasonal shifts in variance. Furthermore, the model revealed strong, seasonally-varying correlations between the species. 
These findings are consistent with an empirical approach and similar to those from the computationally expensive parametric Bayesian Hierarchical Modelling of Species Communities (HMSC) framework.
Consistency, accuracy and feasibility of our approach are demonstrated in a simulation study for up to 10 species.
 }


\begin{document}
\let\thefootnote\relax\footnotetext{* Co-First Authors}

\section{Introduction}

Ecology studies how individual species and community interactions\textemdash{}spanning from mutualism and commensalism to predation and competition\textemdash{}are governed by a range of covariates, such as physical habitat, climatic conditions, and the broader spatiotemporal context.
Species distribution models (SDMs) have become popular tools for describing the relationship of these covariates with the distribution of a species' occurrence or abundance. 
Predictions from SDMs are valuable not only in
fundamental ecological research
but also in applied ecology, especially for the conservation of
endangered species and the management of invasive ones.
Binary regression models and
classification machine learning methods are often used to predict the
presence or absence of a species in a given habitat
\citep[\eg][]{Elith_Leathwick_Hastie_2008,hothorn2011decomposing, cowansSampleSizeConsiderations2025}.
Models describing species abundance rely on count data instead
of presence/absence information and thus yield more
detailed insights into a species' distribution. 

Standard SDMs inherently ignore the interspecific relationships within an ecological community. To address this limitation, and building upon early work in community-level modeling \citep{Death_2002}, more elaborate models that integrate both community and environmental aspects have been proposed over the past decade.
Early ad hoc approaches included incorporating the presence or absence of one species as a
predictor variable in the SDM for another \citep{Kissling_Dormann_Groeneveld_2012}.
Seeking to better disentangle shared environmental preferences from true biotic interactions, the field shifted toward more sophisticated model-based approaches. 
\citet{Ingam_Vukcevic_Golding_2020} categorized these into multi-species distribution models (MSDMs)—which model multiple species simultaneously to share information across their environmental responses \citep[\eg][]{Ovaskainen_Soininen_2011}, sometimes utilizing \textit{a priori} structured correlations to encode phylogenetic information \citep{Ives_Helmus_2011}—and joint species distribution models (JSDMs).

Conversely, JSDMs are defined as statistical frameworks that simultaneously model the joint distribution of multiple species while explicitly accounting for both shared environmental responses and residual correlations between species co-occurrence patterns.
These approaches collectively rely on 
multivariate generalized linear mixed models (MGLMMs) and latent variable models to explicitly capture community dynamics 
\citep[\eg][]{Ovaskainen_Hottola_Siitonen_2010, Pollock_Tingley_Morris_2014, Warton_Blanchet_OHara_2015}. 
While JSDMs typically rely on unstructured residual correlations, recent advancements have also introduced structured correlations to improve the modeling of these interspecific dependencies \citep{Heaps2024}.
Despite these conceptual advancements, the modeling approaches implemented by MSDMs and JSDMs remain largely mutually exclusive \citep[see][for an empirical comparison]{norberg2019comprehensive}:
MSDMs estimate interpretable marginal models for each species but lack an explicit
assessment of their relationships, whereas JSDMs allow the identification of 
species relationships without providing interpretable marginal models.

This motivates a model in which the dependence structure itself varies with covariates rather than being treated as a static residual term.
In the following, we therefore focus on a covariate-dependent copula framework that can retain interpretable marginal SDMs while allowing pairwise associations to change over time or across other environmental gradients.
In econometrics, time-varying copulas are used to estimate the correlation of different asset values depending on time in an autoregressive manner \citep{pattonModellingAsymmetricExchange2006}. A key problem is that they are designed for continuous time series, limiting their applicability to count data and covariates other than time.
Within ecology, studies such as \citet{petren1998habitat} or \citet{bakker2006herbivore} have shown that competition/correlation patterns may also change in response to changes in environmental conditions.
However, most modern JSDMs are MGLMMs or generalized linear latent variable models (GLLVMs),
such as those introduced by \citet{nikuEfficientEstimationGeneralized2019} and \citet{mcgillycuddyParsimoniouslyFittingLarge2025}. 
Neither MGLMMs nor GLLVMs allow for covariate-dependent correlations. 
\citet{huiSpatiotemporalJointSpecies2023} extended MGLMMs by adding a spatiotemporal (spline-based) random effect, where correlation is spatially (and potentially temporally) structured.  
  \citet{hanMaximumLikelihoodEstimation2020} introduced Gaussian copula models for geostatistical count data; 
  and \citet{andersonPathwayMultivariateAnalysis2019} and \citet{popovicGeneralAlgorithmCovariance2018} modeled multivariate correlated counts via a latent Gaussian copula, but none of these approaches allow for covariate-dependent correlations. Furthermore, instead of estimating all parameters jointly, these copula-based methods rely on a two-step procedure---estimating the marginals before the static copula parameters---which leads to incorrect standard errors. Conditional graphical LASSO \citep{augugliaroCglassoPackageConditional2023} estimates the copula and marginals jointly, but is limited to static correlations and cannot accommodate count data.

A relevant extension to MGLMMs was developed by \citet{tikhonovUsingJointSpecies2017} within the Hierarchical Modeling of Species Communities (HMSC) framework, which does allow for covariate-dependent correlations. This model class was identified as the best-performing approach in a large-scale benchmark comparison of 33 JSDMs \citep{norberg2019comprehensive}, making it the state-of-the-art competitor for our method. However, the HMSC implementation is restricted to (log-)Poisson distributed counts and comes with the usual caveats of Bayesian MCMC estimation, such as long computation times and potential convergence issues.

In the following, we present JCTMs, a novel and flexible JSDM framework designed to overcome these limitations by estimating the marginal distributions and covariate-dependent correlations in a data-driven, distribution-free manner. 
Our approach is built upon marginal SDMs and a covariate-dependent copula connecting them. 
Standard SDMs typically rely on rigid parametric distributions (e.g., Poisson or negative binomial) or apply arbitrary \textit{a priori} transformations (\eg $\log(y+1)$) that incorrectly assume continuous likelihoods for discrete data. To resolve this, our marginal SDMs employ the count transformation models of \citet{Siegfried_Hothorn_2020}. This distribution-free framework estimates the data transformation simultaneously with covariate effects by maximizing the exact discrete log-likelihood, naturally accommodating complex features like over- and underdispersion. We extend this foundation to the location-scale transformation models of \citet{siegfriedDistributionFreeLocationScaleRegression2023}, allowing both the location and scale of the transformed counts to depend on covariates.
The structure of the JCTM emerges from
  multivariate conditional transformation models
  \citep{kleinMultivariateConditionalTransformation2022}, in which the correlation
  structure is described by a covariate-dependent latent Gaussian copula.
Joint maximum likelihood estimation and inference for all parameters is performed using the appropriate count likelihood and its gradient 
from~\citet{hothornNonparanormalLikelihoods2024,vign:mvtnorm} implemented in~\citet{pkg:mlt}. 

One could obtain marginal SDMs from commonly applied JSDMs (\ie MGLMMs and GLLVMs) by integrating over the species-random effects, but the simple interpretation of parameters in linear predictors is lost in the process \citep{Lee_Nelder_2004,Muff_Held_Keller_2016}. 
For JCTMs, however, marginal SDMs are directly available as a by-product of the model fitting procedure.

JCTMs are introduced in Section~\ref{sec:methods}.
An example application of an JCTM is provided in Section~\ref{sec:case-study}, comparing it to an empirical approach and HMSC.
The resulting model-based correlation curves and marginal distributions (i.e., SDMs) are presented in Section~\ref{sec:case-study} and discussed in Section~\ref{sec:discussion}, along with potential directions for future work.
A simulation study investigating how feasibility, consistency and accuracy of JCTMs change with sample size and the number of species is provided in the Appendix.
 
\section{Joint count transformation models (JCTMs)} \label{sec:methods} Our approach extends Multivariate Conditional Transformation Models, originally proposed by \citet{kleinMultivariateConditionalTransformation2022}, to multivariate count responses.
JCTMs model the joint distribution of $J$ species count variables $\rYY = (\rY_1, \dots, \rY_J) \in {\NN_0}^J$ conditional on a set of covariates $\rx$. Hereafter, the subscripts $j$ and $\jmath$ refer to any of the $1,\dots,J$ species.

\subsection{General model structure}

We combine the individual SDMs introduced in Section~\ref{sec:marginalmod} into a JSDM leveraging a latent Gaussian copula $\Phi_{\nullvec, \mSigma(\rx)}$, that is, using the cumulative distribution function of a multivariate normal distribution with a zero mean vector and a covariate-dependent correlation matrix $\mSigma(\rx)$ (i.e., constraining $\mSigma(\rx)_{jj} = 1$ in Section~\ref{sec:sigmaX}).
The JCTM is defined by the joint cumulative distribution of $\rYY$ conditional on $\rX\!=\!\rx$ as
\begin{equation}
  \Prob\left(\rYY \leq \yvec \mid \rX\!=\!\rx\right)
= \Phi_{\nullvec, \mSigma(\rx)}\left(\hvec(\yvec \mid \rx)\right),
  \label{eq:general_model}
\end{equation}
where we call
$\hvec(\yvec \mid \rx) = (\h_1(\ry_1 \mid \rx),\dots, \h_J(\ry_J \mid \rx))^\top$ a conditional transformation function.
In order to yield a valid discrete multivariate distribution for count data, $\h_j(\cdot \mid \rx)$ needs to be monotonically increasing (for all $\rx$) and constant on the interval $\left[k-1, k\right)$ for any integer $k\in\NN$ to ensure that the model assigns probabilities only to integer counts.
Furthermore, the transformation of $\ry_j$ may depend on covariates $\rx$, enabling covariate-dependent shifts, scaling, and stratification of the response distribution.

\subsection{Marginal species distribution models} \label{sec:marginalmod}
A key feature of the Gaussian copula framework is that the joint model in Equation~\ref{eq:general_model} decomposes into separate marginal distributions, which serve as the individual SDMs for each species:
\begin{equation} \label{eq:marginal_model}
  \Prob(\rY_j \leq \ry_j\mid\rX\!=\!\rx) = \Phi_{0,1}(\h_j(\ry_j \mid \rx)).
\end{equation}
The conditional probability mass function for an observed count $\ry_j \in \NN_0$ of species $j$ is then
\begin{equation} p_{\rY_j \mid \rX = \rx}(\ry_j)
=
  \begin{cases}
    \Phi_{0,1}(\h_j(\ry_j \mid \rx)) - \Phi_{0,1}(\h_j(\ry_j\!-\!1 \mid\rx)),
    & \text{if } y_j = 1, 2 ,\dots\\
    \Phi_{0,1}(\h_j(\ry_j \mid \rx)), & \text{if } y_j = 0
  \end{cases}.
\end{equation}

Following the location-scale framework of \citet{siegfriedDistributionFreeLocationScaleRegression2023}, we write each conditional transformation function $\h_j(\ry_j \mid \rx)$ in terms of its own baseline transformation $\h_{j}(\ry_j)$, a covariate-dependent shift term $\betavec_{j}(\rx)$, and a covariate-dependent scale term $\gammavec_{j}(\rx)$:

\begin{equation} \label{eq:conditional_h}
  \h_j(\ry_j \mid \rx)= \h_j(\ry_j) \sqrt{\exp(\gammavec_j(\rx))} - \betavec_j(\rx).
\end{equation}
Here, $\gammavec_{j}(\rx)$ and $\betavec_{j}(\rx)$ are modeled as linear predictors $\rx^\top\gammavec_j$ and $\rx^\top\shiftparm_j$.
The baseline transformation function $\h_j(\ry_j)$ generalizes commonly used response transformations such as log- or square-root transformations, by estimating the transformation from the data, similar to but more flexible than Box-Cox transformations \citep{boxAnalysisTransformations1964}.
Following \citet{Siegfried_Hothorn_2020}, we parameterize them as a polynomial in Bernstein form $\avec(\lfloor \ry_j \rfloor)^\top\parm_j$, where $\lfloor \ry_j \rfloor$ is the integer part of $\ry_j$. Monotonicity is ensured using linear constraints on the parameter vector $\parm_j$ as described in \citet{Hothorn_Moest_Buehlmann_2017}.
Note that $\betavec_j(\rx) = \Ex (\h_j(Y_j) \sqrt{\exp(\gammavec_j(\rx))} \mid \rX\!=\!\rx)$ represents the conditional mean of the scaled-transformed counts.

\subsection{Covariate-dependent correlation} \label{sec:sigmaX}
The covariate-dependent correlation matrix $\mSigma(\rx)$ is the central component for modeling interspecific associations. To ensure that $\mSigma(\rx)$ is always a valid correlation matrix during estimation, we parameterize it by means of its scaled inverse Cholesky factor
\begin{equation}
  \mLambda(\rx) =
  \begin{pmatrix}
1 & & & 0 \\
    \lambda_{21}(\rx) & 1 & & \\
    \vdots  & \ddots & \ddots & \\
    \lambda_{J1}(\rx) & \dots & \lambda_{J,J-1}(\rx) & 1
\end{pmatrix},
\end{equation}
where each lower unconstraint off-diagonal element $\lambda_{\jmath j}(\rx)$ (for $\jmath>j$) is expressed as a linear predictor $\rx^\top\xivec_{\jmath j}$.
The correlation matrix $\mSigma(\rx)$ is then computed via:
\begin{gather}
  \mSigma(\rx) = \text{diag}(\mLambda(\rx)^{-1}\mLambda(\rx)^{-\top})^{-1/2}\, \mLambda(\rx)^{-1}\mLambda(\rx)^{-\top} \, \text{diag}(\mLambda(\rx)^{-1}\mLambda(\rx)^{-\top})^{-1/2},
  \label{eq:Sigmax}
\end{gather}
where $\text{diag}(\mathbf{A})^{-1/2}$ denotes the diagonal matrix with the reciprocal inverse square roots of the diagonal elements of a matrix $\mathbf{A}$.
This scaling ensures the diagonal elements of $\mSigma(\rx)$ are always equal to one, yielding a valid correlation matrix.
A detailed discussion of this parameterization and scaling can be found in \citet{hothornNonparanormalLikelihoods2024}.

While the parameters $\lambda_{\jmath j}(\rx)$ have a direct interpretation \citep{Pourahmadi_2007},
we focus on the resulting correlations $\mSigma(\rx)_{\jmath j}$ as they quantify the strength of association between species $\jmath$ and $j$ on the latent transformed scale.
To make the interpretation of those correlations agnostic of the monotone transformation $\hvec(\cdot \mid \rx)$, we can convert them to Spearman's rank correlation coefficient~$\rho$ or Kendall's~$\tau$ \citep{fangMetaellipticalDistributionsGiven2002}:
\begin{align}
  \mSigma^{(\rho)}(\rx)_{\jmath j} &= \frac{6}{\pi} \arcsin\left(\frac{1}{2}\mSigma(\rx)_{\jmath j}\right) \\
  \mSigma^{(\tau)}(\rx)_{\jmath j} &= \frac{2}{\pi} \arcsin\left(\mSigma(\rx)_{\jmath j}\right)
\end{align}

\subsection{Likelihood inference} \label{sec:likelihood_inference}

Count data $\yvec_i \in \NN^J$ can be viewed as interval-censored information
\citep{Siegfried_Hothorn_2020}.
Instead of observing an exact discrete multivariate observation $\yvec_i \in \RR^J$,
a continuous random variable censored to the multidimensional interval $(\yvec_i - \onevec,
\yvec_i]$ is observed.
Thus, for $N$ observations $\yvec_i, i = 1, \dots, N$ of a multidimensional count variable $\rYY = (\rY_1, \dots \rY_J)^\top$
the exact log-likelihood in this framework is given by
\begin{equation} \label{eq:exact_ll}
  \ell(\thetavec) = \sum_{i = 1}^N \log \left(
    \int_{\hvec(\yvec_i - \onevec \mid \rx_i)}^{\hvec(\yvec_i \mid \rx_i)}
  \phi_{\nullvec, \mSigma(\rx_i)} (\zvec) d \zvec  \right)
\end{equation}
where
$\phi_{\nullvec, \mSigma(\rx)}$ is the density of $\Phi_{\nullvec,
\mSigma(\rx)}$ and $\thetavec$ contains all $J$ marginal model parameters $(\gammavec_j, \betavec_j, \parm_j)$ and all covariance parameters of $\mLambda(\rx)$ (\ie $\xivec_{\jmath j}$).
To correctly handle zero counts, we define $\hvec(y \mid \rx_i) = - \infty$ for $y < 0$.
Additionally, we handle missing values in the $j$-th dimension of $\yvec_i$ by setting the integration limits for that dimension in Equation~\ref{eq:exact_ll} to $(-\infty,\infty)$, rendering this dimension for this observation non-informative without ignoring the non-missing entries of $\yvec_i$.

Maximization of the exact log-likelihood $\ell(\thetavec)$ requires fast
numerical procedures for the evaluation of multivariate normal integrals
over hypercubes in the presence of observation-specific, unstructured
covariance matrices $\mSigma(\rx_i)$. A novel implementation of Genz' algorithm
\citep{Genz_1992} in the \pkg{mvtnorm} \proglang{R} add-on package \citep{vign:mvtnorm}
provides functionality for the evaluation of $\ell(\thetavec)$ and the
corresponding gradient when $\mSigma(\rx_i)$ is parameterized in terms
of its Cholesky factor $\text{diag}(\mLambda(\rx_i)^{-1}\mLambda(\rx_i)^{-\top})^{-1/2}\,
\mLambda(\rx_i)^{-1}$ (which is simple to derive from the lower triangular
matrix $\mLambda(\rx_i)$).
As estimating a zero-mean covariate-dependent Gaussian copula is a convex optimization problem \citep{barrattCovariancePredictionConvex2023}, this
allows the application of standard solvers to the
underlying optimization problem
\citep[see][for technical details]{hothornNonparanormalLikelihoods2024}.

The variance-covariance matrix of all model parameters can be obtained from the numerically evaluated Hessian, allowing for Wald-type confidence intervals.
Because we estimate all model parameters jointly---rather than employing a two-step procedure that estimates marginals prior to copula parameters---standard asymptotic theory applies, yielding asymptotically correct standard errors.
Uncertainty in the estimated correlation matrix $\mSigma(\rx)$ can be estimated via a Wald-type parametric bootstrap, where the parameters of $\mLambda(\rx)$ are sampled from their asymptotic distribution.

\section{Case study: Fish-eating aquatic birds} \label{sec:case-study}

\snote{Case study: three competing bird species}
The development of JCTMs was motivated by the species community of three piscivorous aquatic birds, which feed on similarly sized fish within the same habitat: 
Great Cormorant (\emph{Phalacrocorax carbo}), Great Crested Grebe (\emph{Podiceps cristatus}), and Goosander (\emph{Mergus merganser}).
In Europe, these birds can play a major role in controlling the abundance of their prey fish species and, consequently, influence their own respective abundances.

\subsection{Data and study area} \label{sec:data_study_area}
\begin{figure}[htbp]
\begin{knitrout}
\definecolor{shadecolor}{rgb}{0.969, 0.969, 0.969}\color{fgcolor}
\includegraphics[width=\maxwidth]{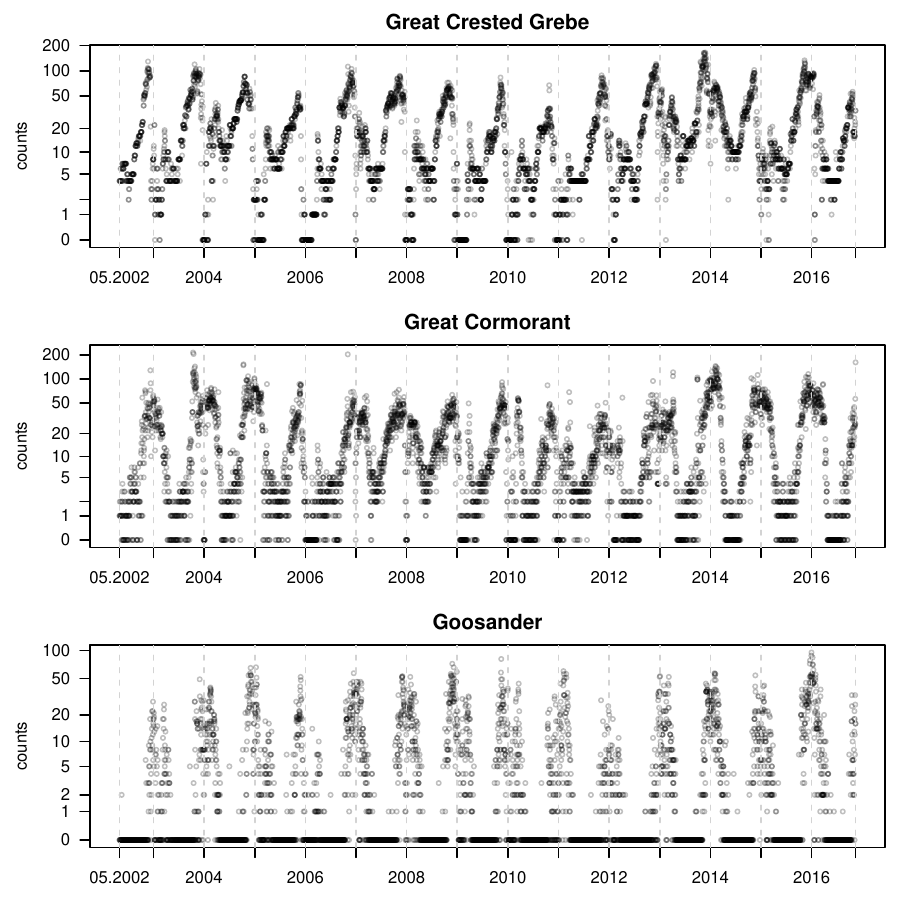} 
\end{knitrout}
  \caption{Observed count time series for the three aquatic bird species at lake Seehamer See (Germany).}
  \label{fig:timeseries}
\end{figure}

The analysis is based on daily
counts of these bird species sampled at lake Seehamer See (Figure~\ref{fig:timeseries}), which
is located in the foothills of the Alps in southern Bavaria, about 40~km
southeast of Munich. It has an area of 1.47~km$^2$, with an average depth of 3.8~m (maximum 12~m).
Thus, the entire water body is accessible
to all three bird species.
\snote{Consistent sampling design ensures data quality}
A local ornithologist, Gerhard Kinshofer, counted
the three species daily between May 1, 2002 and November 13, 2016,
from various locations around the lake using a spotting scope \citep{Kinshofer}. He was therefore
able to spot most, if not all, birds on the lake. As the sampling design remained
identical across all years, it can be safely assumed that any sampling
error was consistent over time.
\snote{Handling of missing data leads to N=4958}
The observation period comprised 5,311 days. Data were missing for 0.34\% of the days for some bird species, when the ornithologist was unsure whether a species was truly absent,
and for 6.31\% of the days for all bird species due to adverse meteorological conditions.
Missing values were handled in the likelihood as described in Section~\ref{sec:likelihood_inference}, implicitly assuming missing completely at random (MCAR).
While adverse weather conditions could potentially affect bird presence or detectability, the overall low missingness rate makes it negligible for demonstrating the JCTM framework.
Thus, we expect any deviations from this assumption to have only a minor influence on our results.

\snote{Data Description}
The raw time series data (Figure~\ref{fig:timeseries}) reveal clear seasonal patterns for all three species, albeit with noticeable year-to-year variation in counts.
All three species showed lower counts during the summer months.
This pattern was particularly pronounced for the Goosander,
with only 13 out of 1,308 observations in June, July, and August being non-zero.
Conversely, the Great Crested Grebe and Great Cormorant exhibited consistent peaks in
counts around November. The rate of decline following this peak
differed noticeably between years, leading to high variation in the subsequent
winter months (December through March).
These complex distributional features---seasonal means, high zero-counts,
and seasonally driven variance---are precisely what our marginal SDMs are designed to capture.

\subsection{Ecological hypotheses and research questions} \label{sec:competition_hypothesis}
\snote{Competition hypothesis from limited food resources}
In the study area, the abundances of all three species in winter increased
during the second half of the 20th century such that carrying capacity was reached
in the 1990s \citep{suter1995cormorants}.
A possible explanation for the logistic growth of the winter populations is
that the food resources for these piscivores were limited.
All three species forage on fish of almost the same size
(10-20~cm) by diving to a depth of 5~m.
Thus, the three species exhibit considerable overlap in their feeding niches.
Accordingly, if common resources are limiting, interspecific competition would be expected.
Furthermore, the associations between the three bird species would presumably change according to seasonal variations in the abundances of the bird species \citep[for seasonal variation of competition see \eg][]{wignall2020seasonal,cecala2020seasonal}.
\snote{Research questions: annual abundance and pairwise correlations}
Hence, in the absence of experimental intervention data, our aim was to answer the following questions:
(1) How does the abundance of each species vary over the course of a year?
(2) How do interspecific associations between the species vary over the course of a year?
That is, what are the Spearman rank correlations between the marginal species abundances, conditional on the day of the year?

\subsection{Modeling bird abundances and associations using JCTMs}\label{sec:apply_JCTM_birds}
Using the JCTMs introduced in Section~\ref{sec:methods}, we analyzed the joint conditional distribution of the trivariate bird-count response $\rYY = (\rY_1, \rY_2, \rY_3)^\top \in {\NN_0}^3$ (1 = Great Crested Grebe, 2 = Great Cormorant, 3 = Goosander), conditional on the day of the year.
Question (1) is addressed by the marginal SDMs (Section~\ref{sec:marginalmod}), and Question (2) is addressed by the covariate-dependent correlations (Section~\ref{sec:sigmaX}).

\snote{Parameterization of the model}
We specified the model components $\h_j(\ry_j)$, $\gammavec_j(\rx)$, $\betavec_j(\rx)$, and $\lambda_{\jmath j}(\rx)$ as follows.
The transformation function $\h_j(\ry_j)$ was parameterized using a polynomial in Bernstein form of order six
as used for count data by \citet{Siegfried_Hothorn_2020}, to significantly reduce the number of parameters while allowing substantial flexibility in the marginal models.
To account for seasonal patterns in abundance and associations, we used a periodic fourth-order Fourier basis of the day-of-the-year as the linear predictor for the shift term $\betavec_j(\rx)$, the scale term $\gammavec_j(\rx)$, and each correlation parameter $\lambda_{\jmath j}(\rx)$.

\begin{figure}[htbp]
\begin{knitrout}
\definecolor{shadecolor}{rgb}{0.969, 0.969, 0.969}\color{fgcolor}
\includegraphics[width=\maxwidth]{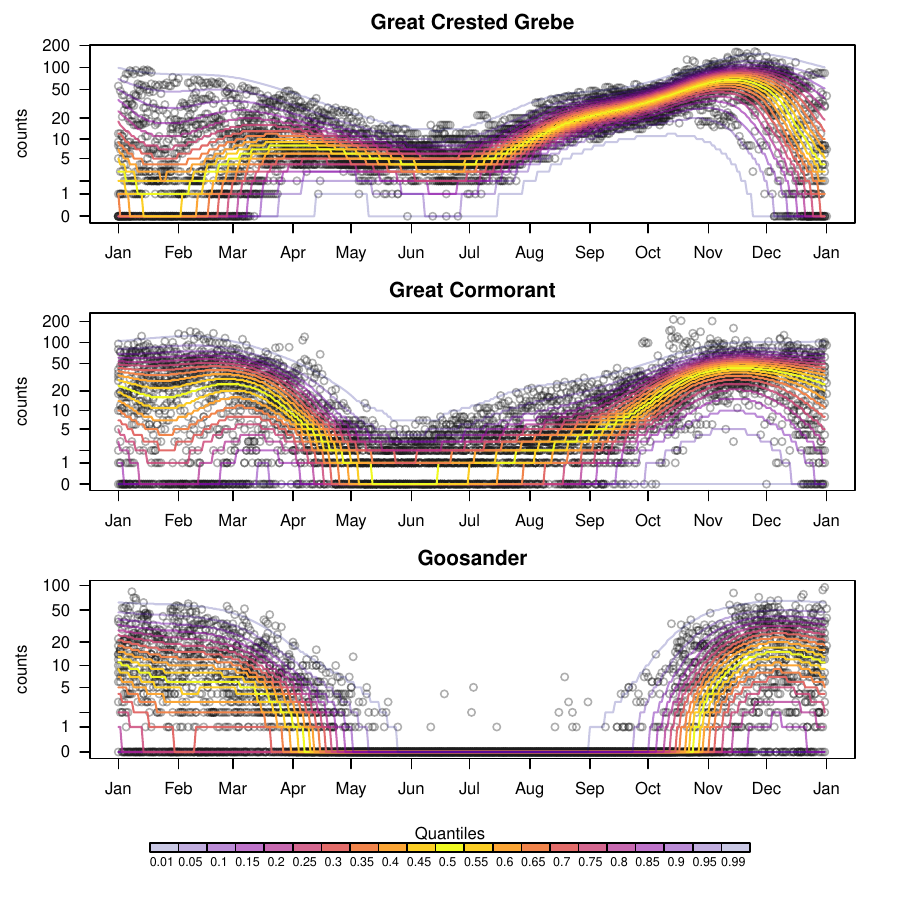} 
\end{knitrout}
  \caption{Estimated marginal distributions conditional on time of the year for the three bird species through our Joint Count Transformation Model (JCTM). Distribution is visualized via quantile curves with observed counts for all years visible as a reference.
}
  \label{fig:marginal_dist}
\end{figure}

\snote{Single SDMs / Marginals}
Figure~\ref{fig:marginal_dist} displays the estimated marginal distributions from our individual SDMs (Equation~\ref{eq:marginal_model}) as quantiles
conditional on the day of the year, addressing our first research
question. The models successfully captured the observed seasonal patterns.
Notably, the inclusion of the covariate-dependent scale term ($\gammavec_j(\rx)$)
allowed the model to account for the higher variability in Great Crested Grebe and
Great Cormorant counts observed after the November peak.
For the Goosander, the flexible baseline transformation $\h_j(\ry_j)$
accurately modeled the high frequency of zero counts during summer while
still providing a good fit to the non-zero counts observed in other seasons.

Our maximum likelihood approach converged in 10 minutes on a single core. The main computational effort
for our method lies in the evaluation of the exact log-likelihood
(Equation~\ref{eq:exact_ll}) and its gradient during optimization.
The results detailing the covariate-dependent interspecific correlations (addressing our second research question) are presented subsequently in Section~\ref{sec:interspecific_associations} and Figure~\ref{fig:triple_cor_plot}.

\subsection{Hierarchical modeling of species communities (HMSC)} \label{sec:hmsc_comparison}
Furthermore, we compared JCTMs to HMSC \citep{ovaskainen2017make}, a state-of-the-art framework for analyzing community data \citep{norberg2019comprehensive}.
The key conceptual difference is that HMSC is a fully parametric approach that assumes a Poisson distribution for the species abundances with a log-link function:
\begin{gather}
  Y_j \sim \operatorname*{Poisson}\left(\exp\left(\rx^\top \tilde\betavec_j + {\eps^{(\text{day})}}_j + {\eps^{(\text{residual})}}_j\right)\right),
\end{gather}
where species associations are captured by the covariance of a multivariate latent random effect at the day-of-the-year level, ${\eps^{(\text{day})}}^\top \sim \ND_J(\nullvec, \tilde\mSigma(\rx))$. An extension allows these associations to depend on covariates \citep{tikhonovUsingJointSpecies2017}. The observation-level random effect in the log-link $\eps^{(\text{residual})} \sim \ND_J(\nullvec, \diag(\sigma_1,\dots,{\sigma}_J))$ addresses residual deviations and potential overdispersion as argued by \citep{Ovaskainen2020}.
Even though we use the same linear predictors for the location and covariance terms as in JCTMs, we expect $\rx^\top \tilde\betavec_j$ and $\tilde\mSigma(\rx)$ to differ, as HMSC only models the conditional means of Poisson distributions.
In contrast, JCTMs learn the shape of the entire conditional distribution from the data, avoiding strong distributional assumptions.

In stark contrast to our approach, the HMSC model was fitted using Bayesian MCMC estimation \citep{pkg:Hmsc} and required
10 days and 5 hours to generate 150,000 MCMC samples (including 100,000 burn-in)
per chain, running 4 chains in parallel on 4 cores. Moreover, despite this immense computational overhead, convergence diagnostics for the covariance parameters indicated poor mixing: the median univariate potential scale reduction factor was 5.53, the multivariate potential scale reduction factor was 115.15, and the effective sample sizes were severely low (median 341, minimum 73) \citep{pkg:coda}.

\subsection{Covariate-dependent interspecific associations} \label{sec:interspecific_associations}

\begin{figure}[htbp]
\begin{knitrout}
\definecolor{shadecolor}{rgb}{0.969, 0.969, 0.969}\color{fgcolor}
\includegraphics[width=\maxwidth]{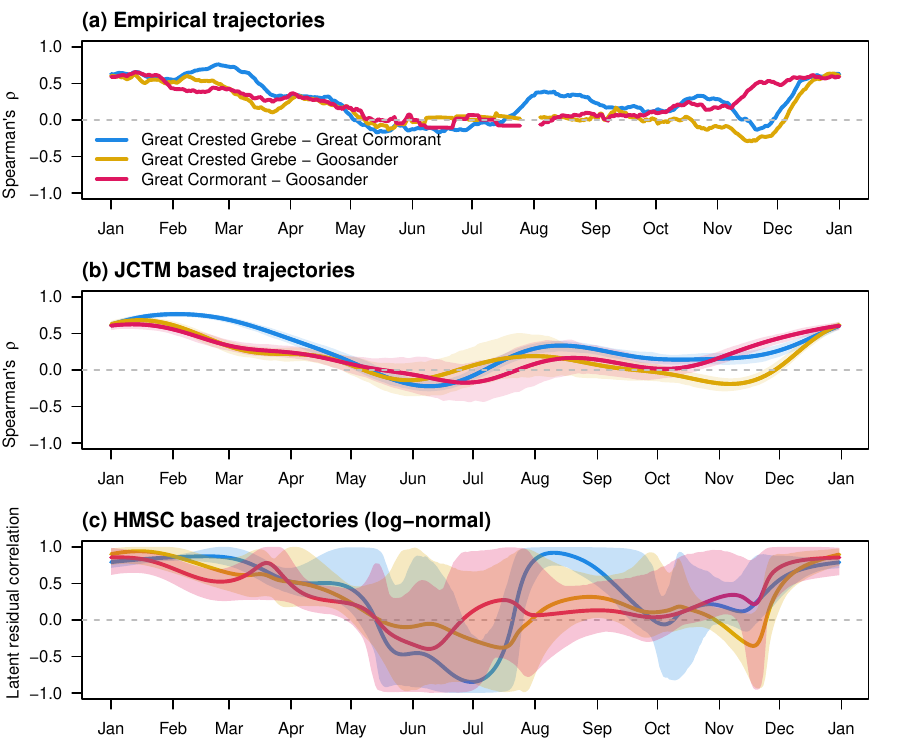} 
\end{knitrout}
  \caption{Pairwise correlation trajectories across species over one year. (a) shows empirically derived Spearman's rank correlations~$\rho$ using all observations within a $\pm 10$ day window across all years; (b) shows JCTM-based Spearman's rank correlations with 95\% confidence intervals derived via parametric bootstrap; (c) shows the corresponding HMSC-based latent residual correlations for the log-normal model. The MCMC chains did not mix/converge well for the covariance parameters, so the posterior-based 95\% confidence intervals should be interpreted with caution. Abbreviations: JCTM, Joint Count Transformation Model; HMSC, Hierarchical Modelling of Species Communities; MCMC, Markov chain Monte Carlo.}
  \label{fig:triple_cor_plot}
\end{figure}

The main motivation for JCTMs was identifying
covariate-dependent interspecies dependencies, addressing our second research question.
Before presenting the two model-based approaches, we first explore these dynamics empirically.
For each day of the year, we computed the rank correlation coefficient (Spearman's $\rho$) for each pair of species using all observations across all years within a $\pm 10$-day window, to obtain a smoother estimate.
Figure~\ref{fig:triple_cor_plot}(a) displays the resulting correlation curves, suggesting higher probabilities of co-occurrences in the colder months and uncorrelated species counts in the warmer months.
\snote{Goal: a model-based dependency estimation}
This empirical window-based approach provides an intuitive initial perspective on the seasonal dependence of interspecific associations. However, the resulting estimates are inherently noisy and lack an uncertainty quantification.

\snote{Sigma(x) + C.I. + empirical comparison + goosander zeros}
The time-dependent Spearman's rank correlations extracted from our model are presented in Figure~\ref{fig:triple_cor_plot}(b).
Our smooth, model-based correlation curves closely resembled the patterns
observed in the empirical correlation trajectories in Figure~\ref{fig:triple_cor_plot}(a).
In particular, the model confirmed the high positive correlations
between all three bird species during the winter months.
We also observed periods of moderate negative correlation, such as between the Great
Crested Grebe and Great Cormorant in June, and between the Great Cormorant
and Goosander in November.
As we observed relevant autocorrelation in the model residuals (not shown), the parametric bootstrap confidence intervals are likely too narrow.
However, we expect them to still give a qualitative impression of the relative uncertainty in the estimates.
And indeed, the uncertainty was
highest for correlations involving the Goosander during the summer months,
a direct consequence of the sparse non-zero counts in that period.
Consequently, we interpreted the correlation estimates in these
high-uncertainty regions with caution.

\snote{HMSC}
For comparison, we fitted the log-normal HMSC variant with covariate-dependent correlations. The resulting correlation curves shown in Figure~\ref{fig:triple_cor_plot}(c) were qualitatively similar to those from our JCTM (Figure~\ref{fig:triple_cor_plot}(b)).
An exact match is not expected, given the conceptual differences:
HMSC models latent residual correlations based on a strict
Poisson assumption for the counts, whereas our model estimates the
rank correlation from flexibly learned, distribution-free marginals.

\section{Discussion}\label{sec:discussion}

JCTMs provide a new perspective on joint
models for species distributions, utilizing flexible distribution-free marginal SDMs combined with a covariate-dependent copula, in a unified maximum likelihood framework for count data.
In this application, we focused on seasonality as the only
covariate of interest influencing both marginal abundances and pairwise correlations.
However, in general, the framework can be extended to include other covariates $\rx$, such as environmental factors,
physical habitat variables, meteorological conditions, and spatiotemporal dynamics.

In contrast to common MSDMs and JSDMs (mostly MGLMMs and GLLVMs),
JCTMs do not make any
assumptions about a parametric distribution of the counts.  Instead, the distribution is estimated in a  semi-parametric way.
This avoids the need to separately consider overdispersion or high zero counts and extends the modeling to situations where common parametric distributions like the Poisson or Negative Binomial are not justified.
Additionally, JCTMs represent the community aspect (\ie measures of dependence) not by a multivariate random residual term in the linear predictor, but rather by pairwise Pearson's correlations between $h_j(y_j\mid\xvec)$ and $h_\jmath(y_\jmath\mid\xvec)$ for species $j$ and $\jmath$, which can be understood as Spearman's rank correlations between species on the original scale of the counts as shown in Figure~\ref{fig:triple_cor_plot} (a,b).
Common MSDMs and JSDMs do not provide such rank-based measures of dependence, but come with Pearson's correlations of random effects on the scale of the link function with no intuitive interpretation.
Moreover, when interpreting fixed effects, JCTMs do not require conditioning on random effects, as is the case when the commonly fixed-/random effects dichotomy is applied.
Consequently, JCTMs allow us to answer a broader class of
research questions that simultaneously involve marginal SDMs and the relationship between
different species depending on covariates $\rx$.
Furthermore, modeling correlations dependent on covariates accommodates
covariate-inflicted variations in
species associations in addition to possible covariate-driven trends in
marginal abundances.

In the relatively simple models for aquatic bird
associations presented herein, the marginal abundances of the three
fish-eating bird species were shown to vary systematically over the course
of a year (hypothesis 1, reflecting different habitat requirements
particularly during the breeding period in summer), with moderate positive
correlations determined in winter and early spring, and small and highly
variable correlations during the rest of the year (hypothesis 2). 
However, as highlighted by \citet{poggiatoInterpretationsJointModeling2021}, residual correlations in phenomenological JSDMs capture a combination of true biotic interactions and shared responses to missing environmental covariates. Therefore, the positive winter associations observed in our model are potentially driven by shared responses to unmeasured environmental factors\textemdash{}such as seasonal fish aggregations or favorable climatic conditions\textemdash{}rather than direct mutualism. Conversely, these strong shared environmental preferences might actually be masking underlying competitive dynamics. 

JCTMs are currently limited in several ways.
With covariate-dependent inverse Cholesky
factor $\mLambda(\rx)$, the model is not as lean on assumptions as a model with constant
correlations, \ie constant parameters in $\mLambda$.
Concretely, the parameter estimates resulting from the joint model might depend
on the order in which the marginal models are specified \citep{barrattCovariancePredictionConvex2023}. This effect can be
counterbalanced by accounting for sufficient flexibility in the marginal models.
Models with covariate-dependent correlations
$\mSigma(\rx)$ should be estimated for different permutations of the
species in a sensitivity analysis.
In general, a strong dependence of the parameter estimates on the order of species indicates a severe lack of model fit.

Alternative interpretations of $\betavec_j(\rx)$, such as odds ratios and hazard ratios, are discussed in \citet{Hothorn_Moest_Buehlmann_2017} and require adapting Equation~\ref{eq:general_model} according to Section 2.6 in \citet{kleinMultivariateConditionalTransformation2022}.
Model estimation is implemented in \pkg{tram} for a range of alternatives.
Stratification can be achieved by estimating the baseline transformation function $h_j$ in Equation~\ref{eq:conditional_h} for each stratum separately. This and more complex interactions between $\rx$ and $\ry_j$ are further discussed in \citet{siegfriedDistributionFreeLocationScaleRegression2023}.
Complex models can be specified for the marginal shift terms
$\betavec_j(\rx)$ and for components of the Cholesky factor $\mLambda(\rx)$, for
example by including either nonlinear or lagging effects of explanatory
environmental variables or terms capturing spatiotemporal trends for
correlated observations.  However, in such cases the likelihood
approach proposed here requires supplementation with appropriate
smoothness constraints.  A mixed version of univariate transformation models
that is also applicable to univariate count transformation models was recently proposed
\citep{Tamasi_2025} and can handle spatially or otherwise correlated
data.  However, the technical challenges associated with mixed
transformation models on the one hand and JCTMs
on the other clearly demonstrate that significant improvements in
this direction will require substantial research efforts.
With respect to
nonlinear habitat effects, a pragmatic approach might be to estimate
marginal SDMs separately for each species, thus allowing nonlinear
associations with environmental variables. The joint models could then be fitted
with the same functional form as that of the marginal shift terms
$\betavec_j(\rx)$.
If the goal is proper statistical inference of model parameters, one should check for conditional independence, for example, by inspecting spatiotemporal residual autocorrelation. The presence of such autocorrelation would indicate that the assumption is violated and reinforces the need to extend the model to a mixed-effects version, as discussed previously.

In a tailored simulation study (see Appendix S1), we evaluated how estimation accuracy and computation time change with the number of species $J$, the sample size $N$, and $M$, the number of Monte Carlo samples used to approximate the log-likelihood in Equation~\eqref{eq:exact_ll}. 
We observed consistent estimation in $N$ and a computational complexity of $O(MN)$, which is expected given the form of the log-likelihood. 
However, the computational burden scales roughly exponentially with $J$. 
This is expected given that the number of parameters in $\mLambda(\rx)$ grows quadratically with $J$, and that more Monte Carlo samples are required to maintain precision when the dimension increases. While computationally challenging, this simulation shows that joint maximum-likelihood estimation for JCTMs is now feasible for communities with up to 10 species, due to the availability of the corresponding score function \citep[see][for computational details]{vign:mvtnorm}. For $J > 15$, there exist several approaches to reduce the computational burden. 
One option is to replace the joint estimation with a two-step procedure, where the marginal SDMs are estimated separately and the copula parameters are estimated in a second step. 
Another option is to treat the counts as a continuous variable and apply the continuous version introduced by \citet{kleinMultivariateConditionalTransformation2022}. However, as our simulation demonstrates, this comes at the severe cost of accuracy and is highly detrimental for sparse, low-count ecological data (such as our Goosander observations).
If the covariate-dependence of the correlation structure is not of interest, one could also consider JSDMs that use a rank-reduction for the correlation structure \citep[\eg][]{mcgillycuddyParsimoniouslyFittingLarge2025}.

JCTMs are potentially complex and cannot be viewed as extensions of
established JSDMs or MSDMs.
Model interpretation requires an understanding of latent, transformed
scales, especially on the marginal scale. Nevertheless, we demonstrated that interesting insights into
the interplay between species can be obtained from a simple
JCTM.
 
\vspace{-.32cm}

\section*{Computational details}

All computations were performed using \proglang{R} version
4.6.0 \citep{R}.
Joint count transformation models are implemented in 
the \pkg{tram} add-on package \citep{pkg:tram}.  
Models and figures from the manuscript and the simulation study can be reproduced with the \pkg{cotram} package \citep{pkg:cotram} using the code below.

\begin{knitrout}
\definecolor{shadecolor}{rgb}{0.969, 0.969, 0.969}\color{fgcolor}\begin{kframe}
\begin{alltt}
\hlkwd{library}\hldef{(}\hlsng{"cotram"}\hldef{)}
\hlkwd{demo}\hldef{(}\hlsng{"aquabirds"}\hldef{,} \hlkwc{package} \hldef{=} \hlsng{"cotram"}\hldef{)}
\hlkwd{source}\hldef{(}\hlkwd{system.file}\hldef{(}\hlsng{'simulation'}\hldef{,} \hlsng{'SIM-JCTM.R'}\hldef{,} \hlkwc{package} \hldef{=} \hlsng{'cotram'}\hldef{))}
\end{alltt}
\end{kframe}
\end{knitrout}

The competitor model HMSC was fitted with the package \pkg{Hmsc} \citep{pkg:Hmsc}.
 
\vspace{-.32cm}
\section*{Acknowledgments}
LG, LB and TH acknowledge financial support by 
Schweizerischer Nationalfonds (grant numbers 200021\_184603 and 200021\_219384).
The authors thank Wendy Ran for improving the language.
We thank the maintainer of the \pkg{Hmsc} package for
support in applying the package to the aquatic bird problem.

\section*{Author Contributions}
LG, LB and TH developed the models and maximum likelihood estimation procedures.
RB contributed the aquatic bird model system. All authors analyzed and
interpreted this data, drafted, and finally revised the manuscript.

\section*{Conflict of Interest Statement} The authors declare no conflicts of interest.
 
\bibliography{birds,packages}

\newpage
\section*{Appendix S1: Simulation study} \label{sec:sim-study}

The empirical case study in Section~3 is intentionally simple, involving only three species.
To assess how JCTMs behave beyond this setting, we conducted a simulation study with increasing dimensionality and sample size.
Our goal is to quantify how estimation accuracy and computation time change with the number of species $J$, the number of observations $N$, and $M$, the number of Monte Carlo samples used to approximate the log-likelihood in Equation~9.
Furthermore, we study how this differs between datasets with low and high counts, as the continuous (later abbreviated as cts.) approximation $\ell_{\text{cts.}}$ from~\cite{kleinMultivariateConditionalTransformation2022} is expected to be less severe for higher counts.
By varying $J$, we assess scalability in higher dimensions; by varying $N$, we assess finite-sample bias and consistency.

\subsection*{Performance metrics}

We evaluate the computation time and the accuracy of the estimated non-transformation parameters $\psivec = (\betavec, \xivec)$, \ie the marginal shift parameters and all covariance parameters of $\mLambda(\rx)$ used for the covariate-dependent correlation structure $\mSigma(\xvec)$.
First, we compute starting values $\hat\psivec_{\text{cts.}}$ using the continuous approximation and record the computation time $t_{\text{cts.}}$ for $M$ samples.
Then, using the Monte Carlo approximation, we compute $\hat\psivec_M$ and record the total computation time $t_M$ in seconds.

To assess the accuracy of the shift parameters, we compute the mean absolute error of estimating $\betavec$:
$$
\operatorname{MAE}_{\betavec} = \operatorname{mean}\left(\left|\betavec - \hat\betavec\right|\right),
$$
For the covariate-dependent correlation matrix, we evaluate the MAE of its off-diagonal elements for $x \in \{0,1\}$, evaluating the error at the boundaries (or endpoints) of the covariate space to capture the maximum potential divergence in the correlation trajectories:
$$
\operatorname{MAE}_{\mSigma} = \operatorname{mean}_{i<j}\left(\left|(\mSigma(0)_{ij}-\hat\mSigma(0)_{ij})\right| + \left|(\mSigma(1)_{ij}-\hat\mSigma(1)_{ij})\right|\right).
$$

To evaluate the accuracy of the Monte Carlo approximation, we compare the log-likelihood $\ell_{\text{M}}$ from Equation~9, approximated using $M$ Monte Carlo samples, with the analogous high-precision baseline $\ell_{10^6}$ computed with $M=10^6$ samples.
To provide a baseline for this approximation, we compare it to the continuous log-likelihood $\ell_{\text{cts.}}$, which ignores the discrete nature of the counts.
For these two comparisons, we inspect the absolute relative differences in log-likelihoods: 
$$
\Delta L_{\text{rel,M}} := \frac{|\ell_{\text{M}} - \ell_{10^6}|}{|\ell_{10^6}|} \quad\text{and}\quad \Delta L_{\text{rel,cts.}} := \frac{|\ell_{\text{cts.}} - \ell_{10^6}|}{|\ell_{10^6}|},
$$
respectively.
Additionally, as a measure of how well the Monte Carlo approximation facilitates the overall optimization, we report the mean absolute gradient of the log-likelihood for $\psivec$, evaluated using $M$ Monte Carlo samples:
$$
\operatorname{MAG}_{M, \psivec} = \operatorname{mean}\left(\left|\nabla_{\psivec} \ell_{\text{M}}(\psivec)\right|\right).
$$
In particular, we evaluate the three variants: $\operatorname{MAG}_{M, \hat\psivec_M}$ capturing how well the optimizer managed to find the optimum of $\ell_M$, $\operatorname{MAG}_{10^6, \hat\psivec_M}$ capturing how well the Monte Carlo approximation performes in comparison to a high precision baseline, and $\operatorname{MAG}_{10^6, \hat\psivec_{\text{cts.}}}$ capturing how well the continuous approximation performs in comparison to the same high-precision baseline.

\subsection*{Data generating process}

As in Section~2.4 we treat $Y$ as a censored version of $Y^\star\in \RR^J$, for which we set up a parametric model. We sample from this model and obtain the corresponding counts by applying the ceiling function to each dimension, $Y = \lceil Y^\star \rceil$.
To keep the setting simple we omit the scale term (\ie $\gammavec_{j}(x) = 1$) and restrict the simulation to a single covariate ${\xvec} \sim \UD([0,1]^N)$, influencing the covariate-dependent correlation and the $j$-th marginal shift effects $\betavec_j(x)=\beta_jx$, where $\betavec \sim \UD([-1,1]^J)$.
For the transformation function $h_j(y)$ we use the inverse of a chi-square cdf with $\df_j$ degrees of freedom.
For the low-count datasets, we sample $\mathbf{\df} \sim \UD(\{3,4,\dots,9\}^J)$, whereas for the high-count datasets we sample $\mathbf{\df} \sim \UD(\{30,31,\dots,50\}^J)$.
For the covariate-dependent correlation we define $\mSigma_0, \mSigma_1$ to be standardized independent samples from a Wishart distribution with an identity scale matrix and $J_{\max}$ degrees of freedom, and $\mLambda_0, \mLambda_1$ their respective scaled inverse Cholesky factors as in Equation~6.
The desired $\mSigma(x)$ is then computed via Equation~6 using $\mLambda(x) = \mLambda_0 + x(\mLambda_1 - \mLambda_0)$.
The above quantities are sampled once and then kept fixed for all simulation scenarios and repetitions; for a scenario with $J < J_{\max}$ we use the leading $J\times J$ submatrices.
Sampling the $i$-th observation $Y^{(i)}$ of a new dataset thus requires sampling $\mU \sim \UD([0,1]^J)$, and computing $Y^{(i)}_j = \lceil F_{Y^\star}^{-1}(\mU)_j\rceil  =  \lceil h_j^{-1}(\Phi_{\nullvec,\Sigma(x_i)}^{-1}(\mU)_j + \beta_jx_i)\rceil$.
In this manner, we generate 15 datasets with $J_{\max}=15$ and $N_{\max}=10000$, and later take the subset containing the first $N$ observations and the first $J$ species for each scenario.
For the low-count datasets, the 0\%, 25\%, 50\%, 75\%, and 100\% quantiles across all datasets are 1, 3, 6, 9, and 49, respectively.
For the high-count datasets, these quantiles are 7, 33, 41, 48, and 112, respectively.

\subsection*{Results and discussion}
To investigate these parameters, we evaluated the scenarios detailed in the figures below.
Each data point in these plots corresponds to the median over 15 repetitions of the simulation, where we computed the aforementioned performance measures and computation times for each repetition.

\begin{figure}[htbp]
\centering
\begin{knitrout}
\definecolor{shadecolor}{rgb}{0.969, 0.969, 0.969}\color{fgcolor}
\includegraphics[width=\maxwidth]{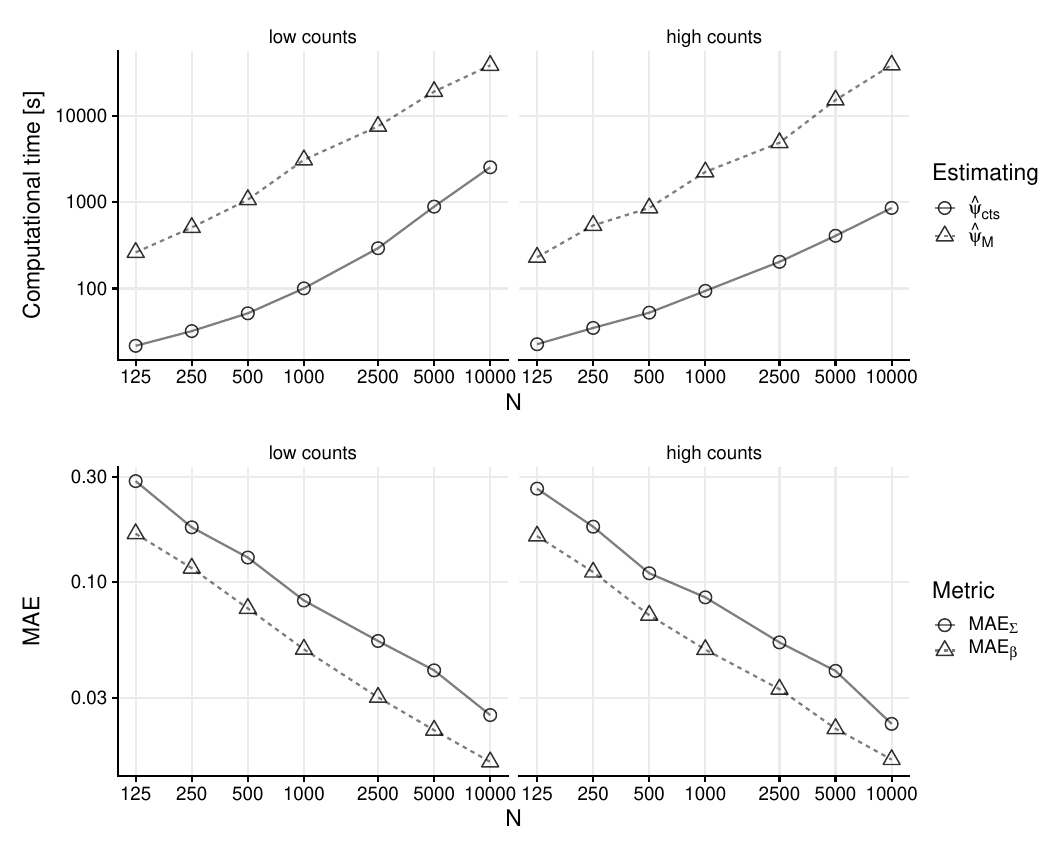} 
\end{knitrout}
\caption{Consistency with increasing sample size $N$ (fixed $J=10$, $M=500$).}
\label{fig:simN}
\end{figure}

\begin{figure}[htbp]
\centering
\begin{knitrout}
\definecolor{shadecolor}{rgb}{0.969, 0.969, 0.969}\color{fgcolor}
\includegraphics[width=\maxwidth]{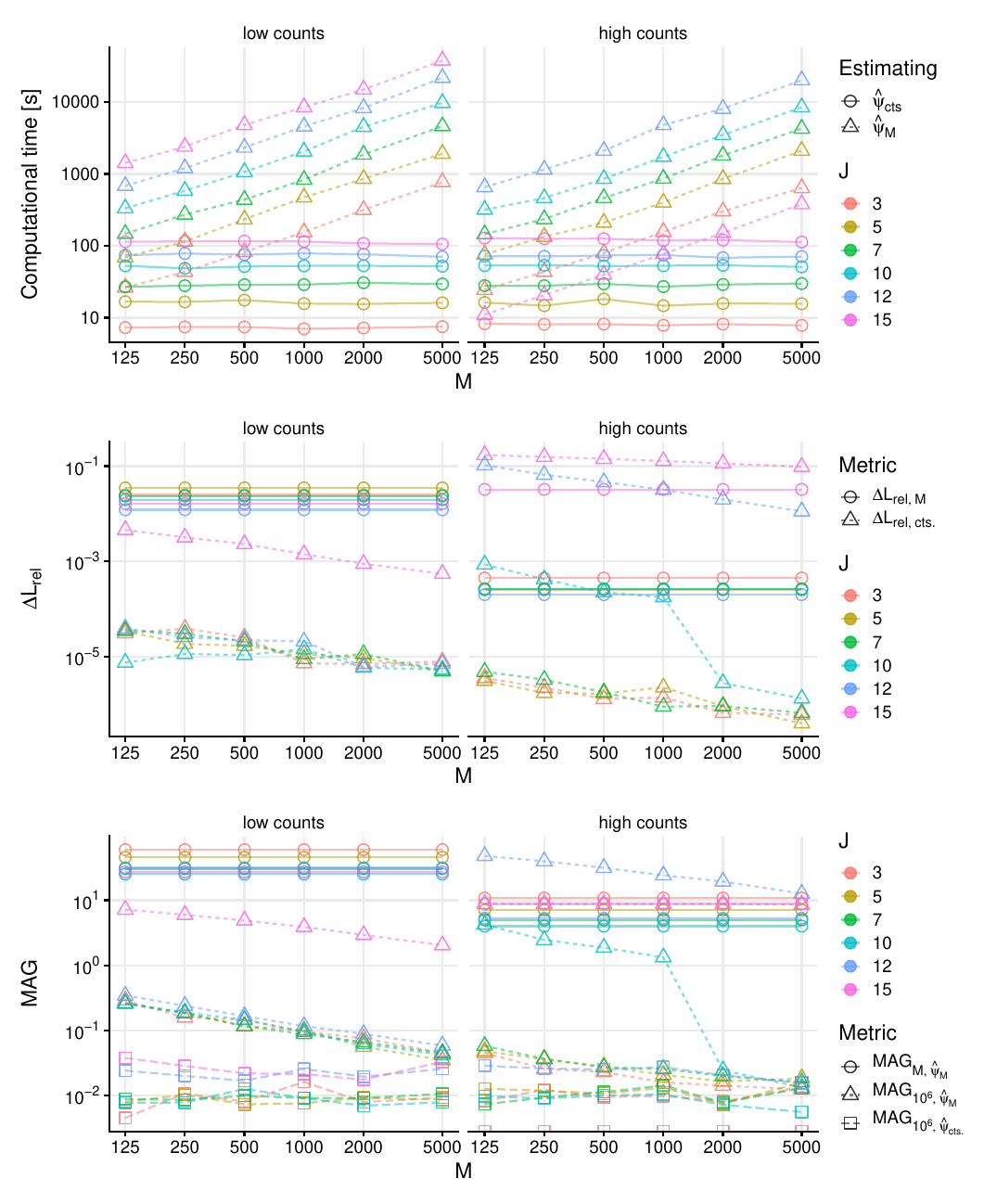} 
\end{knitrout}
\caption{Accuracy of the Monte Carlo approximation with increasing number of samples $M$ and number of species $J$ (fixed $N=500$). \textit{Note: Sudden drops in computation time ($t_M$) or anomalous gradient trajectories for high-count datasets with $J \geq 12$ indicate optimizer convergence failures.}}
\label{fig:simM}
\end{figure}

\begin{figure}[htbp]
\centering
\begin{knitrout}
\definecolor{shadecolor}{rgb}{0.969, 0.969, 0.969}\color{fgcolor}
\includegraphics[width=\maxwidth]{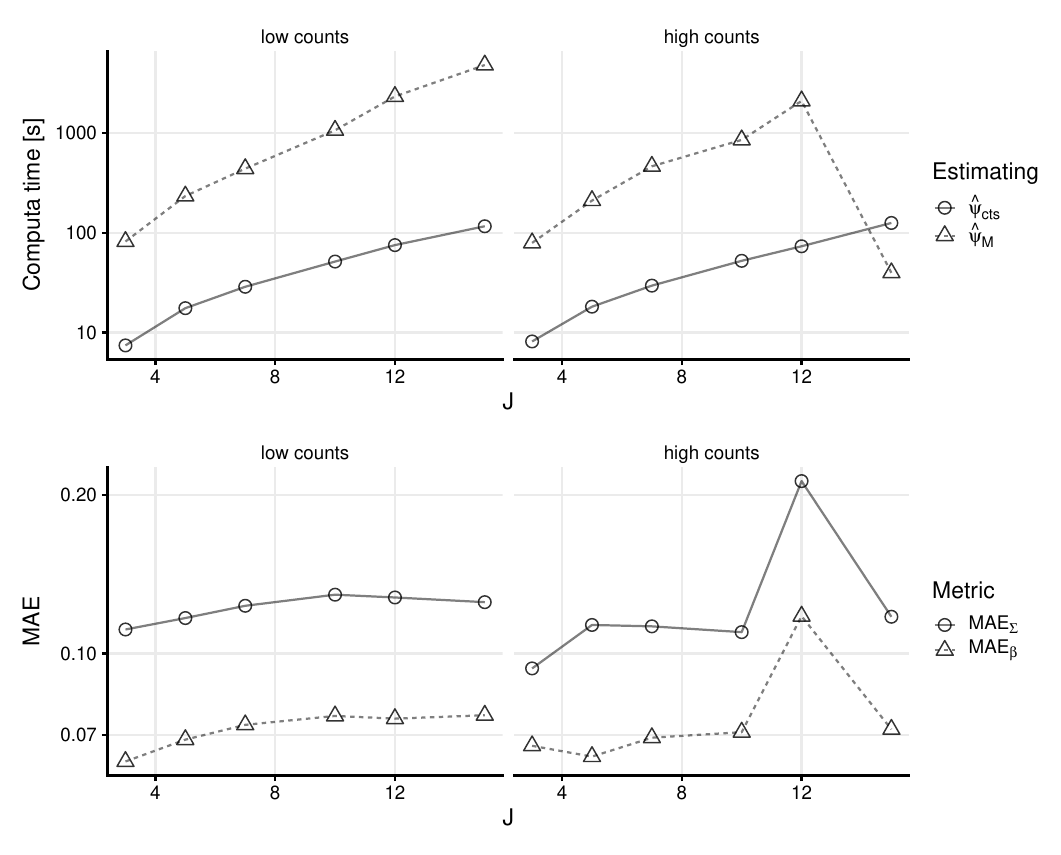} 
\end{knitrout}
\caption{Scalability with increasing number of species $J$ (fixed $N=500$, $M=500$). \textit{Note: Anomalies in computation time and accuracy metrics for high-count datasets with $J \geq 12$ indicate optimizer convergence failures.}}
\label{fig:simJ}
\end{figure}

\paragraph{Consistency with increasing sample size (Figure~\ref{fig:simN}).}
Both the accuracy of the estimated shift parameters ($\operatorname{MAE}_{\betavec}$) and the correlation structure ($\operatorname{MAE}_{\mSigma}$) improve consistently with increasing $N$ for both low- and high-count settings, empirically confirming that the implementation of the theoretically consistent joint maximum likelihood estimator converges to the unknown model parameters.
Computation time scales linearly with $N$, as expected from the form of the log-likelihood in Equation~9.

\paragraph{Accuracy of the Monte Carlo approximation (Figure~\ref{fig:simM}).}
Computation time scales linearly with $M$, as expected.
The relative log-likelihood difference $\Delta\ell_{\text{rel,M}}$ and the mean absolute gradient $\operatorname{MAG}_{10^6, \hat\psivec_M}$ both decrease with increasing $M$, confirming that the Monte Carlo approximation improves with more samples.
Across all settings, the Monte Carlo approximation is substantially more accurate than the continuous approximation $\ell_{\text{cts.}}$, with the gap being particularly pronounced for low-count data, where treating discrete counts as continuous is least appropriate. 
Furthermore, comparing the trajectories for $\operatorname{MAG}_{10^6, \hat{\psivec}_M}$ and $\operatorname{MAG}_{10^6, \hat{\psivec}_{\text{cts.}}}$ in Figure~\ref{fig:simM} confirms that the continuous approximation is less severe for high-count data.
This highlights the importance of using the exact discrete likelihood for count data, and demonstrates that the Monte Carlo approximation can achieve this at a manageable computational cost.
The trajectory for $\operatorname{MAG}_{10^6, \hat{\psivec}_M}$ shows that optimization of $\ell_M$ was successful, and this quantity could be further reduced by adapting corresponding convergence tolerance parameters at the cost of increased computation time.

\paragraph{Scalability with increasing number of species (Figure~\ref{fig:simJ}).}
Even when holding $M$ constant, the computational burden scales roughly exponentially with $J$.
This exponential scaling becomes even more apparent if $M$ must be increased to maintain precision in higher dimensions.

We conclude that joint maximum likelihood estimation of JCTMs is reasonably fast and accurate for $J \leq 10$.
For $J \geq 15$, the computational burden quickly becomes prohibitive and the accuracy decreases. Note that for high-count datasets with $J \geq 12$ (Figures~\ref{fig:simM} and \ref{fig:simJ}), the sudden drops in computation time and anomalies in the accuracy metrics indicate numerical underflow and optimizer convergence failures, rather than true efficiency gains. Alternatives such as a two-step procedure---estimating marginal SDMs first and the copula parameters second---or the continuous approximation $\ell_{\text{cts.}}$ may be the only practical options, at the cost of less accurate standard errors or a less appropriate likelihood, respectively.
For applications where covariate-dependent correlations are not the primary interest, dimension reduction approaches such as those proposed by \citet{mcgillycuddyParsimoniouslyFittingLarge2025} offer an additional avenue for scaling to larger communities.
 
\end{document}